\documentclass[twoside,twocolumn,english,aps,showpacs,preprintnumbers]{revtex4}
\usepackage[T1]{fontenc}
\usepackage[latin9]{inputenc}
\usepackage{color}
\usepackage{amsmath}
\usepackage{graphicx}
\usepackage{amssymb}

\makeatletter
\@ifundefined{textcolor}{}
{%
 \definecolor{BLACK}{gray}{0}
 \definecolor{WHITE}{gray}{1}
 \definecolor{RED}{rgb}{1,0,0}
 \definecolor{GREEN}{rgb}{0,1,0}
 \definecolor{BLUE}{rgb}{0,0,1}
 \definecolor{CYAN}{cmyk}{1,0,0,0}
 \definecolor{MAGENTA}{cmyk}{0,1,0,0}
 \definecolor{YELLOW}{cmyk}{0,0,1,0}
 }

\@ifundefined{definecolor}
 {\usepackage{color}}{}
\makeatother

\makeatother

\begin{document}

\title{Quantum phases of Bose-Bose mixtures on a triangular lattice}

\author{Liang He$^{1}$, Yongqiang Li$^{1,3}$, Ehud Altman$^{2}$ and Walter
Hofstetter$^{1}$}

\affiliation{$^{1}$Institut für Theoretische Physik, Goethe\textendash{}Universität,
60438 Frankfurt/Main, Germany }

\affiliation{$^{2}$Department of Condensed Matter Physics, The Weizmann Institute
of Science, Rehovot 76100, Israel}

\affiliation{$^{3}$Department of Physics, National University of Defense Technology,
Changsha 410073, P. R. China }
\begin{abstract}
We investigate the zero temperature quantum phases of a Bose-Bose
mixture on a triangular lattice using Bosonic Dynamical Mean Field
Theory (BDMFT). We consider the case of total filling one where geometric
frustration arises for asymmetric hopping. We map out a rich ground
state phase diagram including\textcolor{red}{{} }$xy$-ferromagnetic,
spin-density wave, superfluid, and supersolid phases. In particular,
we identify a stripe spin-density wave phase for highly asymmetric
hopping. On top of the spin-density wave, we find that the system
generically shows weak charge (particle) density wave order. 
\end{abstract}


\maketitle

\section{Introduction}

Geometric frustration arises when magnetic interactions between different
spins on a lattice are incompatible with the underlying crystal geometry.
Since the first investigation of the Ising antiferromagnet on the
triangular lattice \cite{Ising_ant_ferro_tri}, geometric frustration
has been a constant source of surprises that inspired the development
of new concepts \cite{geo_fru_1,geo_fru_2}. This provides strong
motivation to study the physics of frustration from the different
perspective provided by systems of ultra-cold atoms.

Recent experiments have made substantial strides in this direction
with the realization of non-bipartite triangular \cite{Tri_lat_exp,cla_fru_mag_optlat_Sengstock_2011}
and kagome \cite{Kagome_lat_exp} optical lattices. In this paper
we investigate the consequence of bringing a two component bosonic
mixture into the Mott insulating regime on the triangular lattice.
Deep in the Mott state, the magnetic exchange interactions can include
as the main source of frustration a strong Ising antiferromagnetic
exchange.

Our goal is to study how the magnetic phases evolve when we increase
the frustrating Ising interaction or approach closer to the transition
to the superfluid phase, such that the effective spin model with second-order
exchange interactions no longer applies.

To answer this question we apply Bosonic Dynamical Mean Field Theory
(BDMFT) \cite{BDMFT_1,BDMFT_2,BDMFT_3}, which is non-perturbative
in the hopping amplitudes. We find that even in the deep Mott regime,
the standard spin exchange model is invalid for extremely asymmetric
hopping. Instead, we numerically identify a stripe spin-density wave
(SDW) phase (see Fig.~\ref{Flo:pd_Vb_Vd_48U}a.2). This can be understood
within a higher-order effective spin-model (\ref{eq:H_eff_BFK}) derived
from fourth order perturbation theory. The effective description shows
that the stripe SDW is favored by the higher-order density fluctuations
of the {}``lighter'' atoms which remove the Ising-type frustration.
Moreover, we also find that on top of the spin-density wave, due to
asymmetry of the hopping amplitudes, the system develops a weak charge-density
wave in the total particle density (see Fig.~\ref{Flo:weak_CDW}).

The paper is organized as follows. In Sec. II, we introduce the system
and model studied here, as well as the theoretical approach used in
our investigation. In Sec. III, the main part of this paper, we present
a detailed discussion of the ground state properties of the system.
We conclude in Sec. IV.

\section{Model and method}

We consider two species (hyperfine states or isotopes) of ultracold
bosons loaded into a triangular optical lattice.\textcolor{red}{{}
}For sufficiently low filling this system can be described by a two-component
Bose-Hubbard model in the lowest band approximation:\begin{eqnarray}
H & = & -\sum_{\langle i,j\rangle}(t_{a}a_{i}^{\dagger}a_{j}+t_{b}b_{i}^{\dagger}b_{j}+\textnormal{h}.\textnormal{c}.)+U\sum_{i}n_{ai}n_{bi}\nonumber \\
 &  & +\frac{1}{2}\sum_{i;\alpha=a,b}V_{\alpha}n_{i\alpha}(n_{i\alpha}-1)-\sum_{i;\alpha=a,b}\mu_{\alpha}n_{\alpha i}\,.\label{eq:Hamiltonian}\end{eqnarray}
 Here $\langle i,j\rangle$ denotes nearest-neighbor sites, $a_{i}$($a_{i}^{\dagger}$),
$b_{i}$($b_{i}^{\dagger}$) are bosonic annihilation (creation) operators
of the two species on site $i$ in the Wannier representation, and
$n_{ai}\equiv a_{i}^{\dagger}a_{i}$, $n_{bi}\equiv b_{i}^{\dagger}b_{i}$.
The first term in Eq.~(\ref{eq:Hamiltonian}) describes the kinetic
energy of each species with hopping amplitudes $t_{a}$ and $t_{b}$;\textcolor{red}{{}
}the second and the third term represent the on-site inter-species
interaction $U$ and the intra-species interactions $V_{a}$ and $V_{b}$
for species $a$ and $b$, respectively; finally, $\mu_{a}$ and $\mu_{b}$
denote the chemical potentials.\textcolor{red}{{} }

Previous studies on two-component ultracold bosons in a square or
a cubic optical lattice have revealed a rich phase diagram\textcolor{blue}{{}
}\cite{TBHM_Ehud}. Within the Mott insulator, at low temperature
quantum magnetism arises, in particular $z$-antiferromagnetic and
$xy$-ferromagnetic order. For total filling one, the emergence of
magnetic order can be easily understood if we note that in the deep
Mott regime (strong coupling limit) $t_{a,b}\ll U,V_{a,b}$, the physics
of the two-component Bose-Hubbard Hamiltonian (\ref{eq:Hamiltonian})
is given by\textcolor{red}{{} }an effective spin-$1/2$ $XXZ$ model
\cite{TBHM_Kuklov,TBHM_Duan,TBHM_Ehud} \begin{equation}
H_{\mathrm{eff}}=J_{z}\sum_{\langle ij\rangle}S_{i}^{z}S_{j}^{z}-J_{\perp}\sum_{\langle ij\rangle}(S_{i}^{x}S_{j}^{x}+S_{i}^{y}S_{j}^{y})-h\sum_{i}S_{i}^{z}\label{eq:H_eff}\end{equation}
 where $\mathbf{S}_{i}\equiv(a_{i}^{\dagger},b_{i}^{\dagger})(\boldsymbol{\sigma}/2)\left(\begin{array}{c}
a_{i}\\
b_{i}\end{array}\right)$ with $\sigma_{x},\sigma_{y},\sigma_{z}$ being the Pauli matrices
and \begin{eqnarray}
J_{z} & = & 2\frac{t_{b}^{2}+t_{a}^{2}}{U}-\frac{4t_{a}^{2}}{V_{a}}-\frac{4t_{b}^{2}}{V_{b}},\label{eq:eff_Jz}\\
J_{\perp} & = & \frac{4t_{a}t_{b}}{U},\label{eq:eff_J_perp}\\
h & = & \frac{2t_{a}^{2}}{V_{a}}-\frac{2t_{b}^{2}}{V_{b}}+(\mu_{a}-\mu_{b}).\label{eq:eff_h_magnetic_field}\end{eqnarray}
\textcolor{black}{{} In the following discussion, we assume $h=0$,
i.e. vanishing spin imbalance (in the case of asymmetric hopping $t_{a}\neq t_{b}$,
the chemical potentials $\mu_{a}$ and $\mu_{b}$ are tuned to achieve
this).} On bipartite lattices (e.g. square, cubic) the system supports
$xy$-ferromagnetic order for $J_{\perp}>J_{z}>0$, which is characterized
by the local correlator $\langle a^{\dagger}b\rangle$, and $z$-antiferromagnetic
order for $J_{z}>J_{\perp}>0$, characterized by the order parameter
$\Delta_{\mathrm{af}}=|\langle S_{\alpha}^{z}\rangle-\langle S_{\bar{\alpha}}^{z}\rangle|$
with $\alpha$ ($\alpha=-\bar{\alpha}$) being the sublattice index.

We expect new interesting physics to emerge in a system with triangular
instead of bipartite optical lattice. For antiferromagnetic exchange
coupling $J_{z}$, it is impossible to minimize the energy of the
spin configuration on each lattice bond, i.e. geometric frustration
arises and the system may develop exotic phases at low temperature.
We are particularly interested in the extremely asymmetric hopping
regime for the full range of couplings from Mott insulator to superfluid.
Since higher order density fluctuations of the {}``lighter'' species
are expected to become even larger than the low-order density fluctuation
of the {}``heavier'' ones, their interplay with Ising-type frustration
in $z$ allows the system to form novel phases. Also frustration is
expected to grow with increasing values of the hopping amplitudes
since the exchange coupling arises from the itinerancy of atoms. In
the following, we shall address these two aspects by mapping out the
ground state phase diagram for total filling one per site via BDMFT.

Before going into a detailed discussion of the system's properties,
let us at this point briefly introduce Bosonic Dynamical Mean Field
Theory (BDMFT) \cite{BDMFT_1,BDMFT_2,BDMFT_3}, which is an extension
of Dynamical Mean Field Theory (DMFT) \cite{DMFT_d_inf,DMFT}, originally
developed to treat strongly correlated fermionic systems. It is non-perturbative,
captures local quantum fluctuations exactly and becomes exact in the
infinite-dimensional limit. Note that for antiferromagnetic exchange
$J_{z}$ the ground state of the system may break the translational
symmetry of the lattice. We investigate this system using real-space
BDMFT (R-BDMFT) \cite{R-BDMFT}, which is a generalization of BDMFT
to a position-dependent self-energy and captures inhomogeneous quantum
phases. Within BDMFT/R-BDMFT, the physics on each lattice site is
determined from a local effective action which is obtained by integrating
out all the other degrees of freedom in the lattice model, excluding
the lattice site considered. The local effective action is then represented
by an Anderson impurity model \cite{BDMFT_1,BDMFT_2,BDMFT_3}. We
use Exact Diagonalization (ED) \cite{ED_solver_1,ED_solver_2} of
the effective Anderson Hamiltonian with a finite number of bath orbitals
to solve the local action ($n_{\mathbf{\textnormal{bath}}}=4$ bath
orbitals are chosen in the current work).\textcolor{blue}{{} }Details
of the R-BDMFT method have been published previously \cite{R-BDMFT}.

\section{Results}

Since we are mainly interested in the effects of geometric frustration,
we focus on the parameter regime $V_{a}=V_{b}\gg U$, where the leading
exchange couplings become Ising antiferromagnetic for highly asymmetric
hopping, resulting in Ising-type frustration. In our simulations,
the chemical potentials $\mu_{a}$ and $\mu_{b}$ are tuned to equal
particle number of both species ($N_{a}\equiv\sum_{i}\langle n_{ia}\rangle=N_{b}\equiv\sum_{i}\langle n_{ib}\rangle$)
and a total filling factor $\rho=\sum_{i}\langle n_{ia}+n_{ib}\rangle/N_{\mathrm{lat}}=1$
($N_{\mathrm{lat}}$ is the number of lattice sites.). The full range
of interactions from strong coupling, deep in the Mott phase, all
the way to the superfluid at weak coupling is investigated within
BDMFT.

\begin{figure}
\includegraphics[clip,width=3.3in]{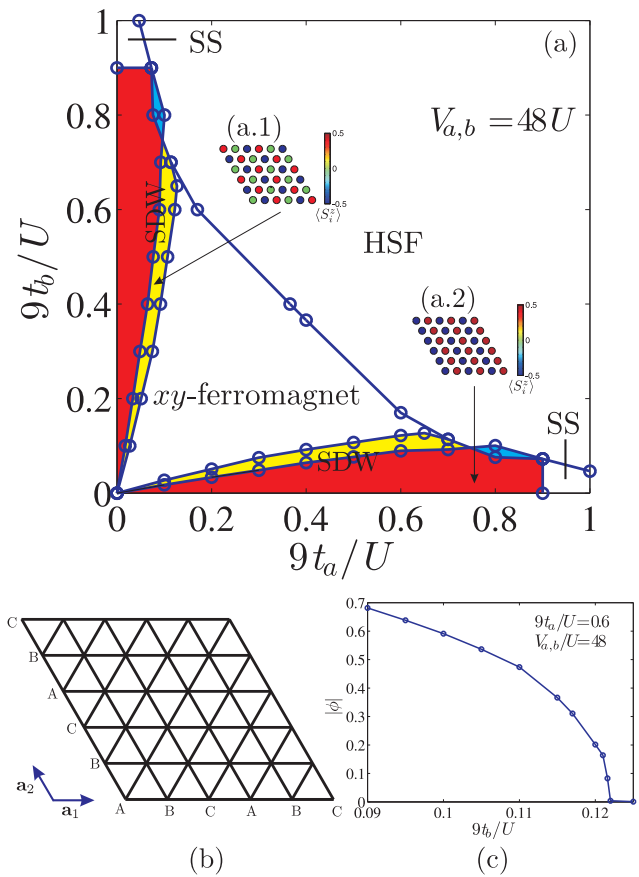}

\caption{(Color online) (a) Ground state phase diagram at unit filling $\rho=1$,
with $V_{a}=V_{b}=48U$, obtained from calculations on a $6\times6$
triangular lattice. We observe four major different phases, which
are the homogeneous superfluid (HSF), $xy$-ferromagnet, spin-density
wave (SDW), and supersolid (SS). In the SDW region, the area where
the SDW has a 3-sublattice structure is marked in yellow, while the
area featuring the stripe SDW is marked red. The insets (a.1) and
(a.2) show the corresponding $z$-magnetization distribution $\langle S_{i}^{z}\rangle$.
The small blue area indicates a coexistence region of HSF and stripe
SDW phases. (b) The triangular lattice considered here with sublattice
indices $A,B,C$ marked near the corresponding lattice sites. $\mathbf{a}_{1}$
and $\mathbf{a}_{2}$ are the two lattice vectors. (c) $t_{b}$ dependence\textcolor{red}{{}
}of the 3-sublattice SDW order parameter at fixed $9t_{a}/U=0.6$. }

\label{Flo:pd_Vb_Vd_48U} 
\end{figure}

Our main results are summarized in Fig.~\ref{Flo:pd_Vb_Vd_48U}a,
which shows the ground state phase diagram for large intra-species
interaction strengths ($V_{a,b}/U=48$). Two different magnetic phases
are found in the Mott insulator. When $t_{a}$ and $t_{b}$ are of
comparable magnitude, the leading exchange coupling is ferromagnetic
in the $xy$-plane, and the system is in the $xy$-ferromagnetic phase,
characterized by a uniform magnetization in the $xy$-plane. On the
other hand, for sufficiently large asymmetry between the two hopping
amplitudes, we observe two types of spin-density wave (SDW) phases
(see insets (a.1) and (a.2) in Fig.~\ref{Flo:pd_Vb_Vd_48U}a), which
break the translation symmetry of the lattice. All of these spin-ordered
phases are found to persist up to the superfluid transition. For large
hopping the ground state breaks $U(1)$ symmetry and develops superfluid
order $\langle a\rangle$,$\langle b\rangle$. Depending on the relative
magnitude of $t_{a}$ and $t_{b}$, the system can also exhibit additional
charge-density wave order in each species, leading to a supersolid.
In the following subsections we discuss detailed properties of the
different phases.

\subsection{Spin-density wave (SDW) in the Mott-insulator region}

Let us now discuss the origin of the two different SDW phases found
in the asymmetric hopping regime.

\subsubsection{{}``3-sublattice'' SDW}

In the less asymmetric hopping regime of the SDW phase region (yellow
area in the phase diagram Fig.~\ref{Flo:pd_Vb_Vd_48U}a), the SDW
features a 3-sublattice structure (see Fig.~\ref{Flo:distr_magnetization}b)
ordering at the wave vectors $\pm\mathbf{Q}=(\pm4\pi/3,0)$. The emergence
of this type of SDW can be understood from a spin-wave theory of the
spin-$1/2$ $XXZ$ model (\ref{eq:H_eff}), which shows that the uniform
$xy$-ferromagnetic phase develops an instability at $\pm\mathbf{Q}=(\pm4\pi/3,0)$
towards 3-sublattice ordering with increasing $J_{z}/J_{\perp}$ \cite{MF_GS_tri_XXZ,Melko_Balents_tri_XXZ}.
Here, we choose the order parameter of the 3-sublattice SDW phase
as \begin{equation}
\phi\equiv\langle S_{A}^{z}\rangle+\langle S_{B}^{z}\rangle e^{i2\pi/3}+\langle S_{C}^{z}\rangle e^{i4\pi/3}\label{eq:3_sub_lat_SDW_order}\end{equation}
 where $\langle S_{A,B,C}^{z}\rangle$ is the $z$-magnetization on
the sites of the $A,B,$ and $C$ sublattice respectively. $\phi$
is the Fourier transform of the $z$-magnetization distribution $\langle S_{i}^{z}\rangle$
at the wave vector $\mathbf{Q}$. In Fig.~\ref{Flo:pd_Vb_Vd_48U}b,
at fixed $t_{a}$, the $t_{b}$ dependence of $|\phi|$ is shown,
indicating a second order phase transition from the 3-sublattice SDW
to an $xy$-ferromagnet.

Let us mention that previous investigations of the spin-$1/2$ $XXZ$
model in the large $J_{z}/J_{\perp}$ region have revealed that the
$z$-magnetization favors a different 3-sublattice patten of $(\langle S_{A}^{z}\rangle=\pm2m,\langle S_{B}^{z}\rangle=\mp m,\langle S_{C}^{z}\rangle=\mp m)$
in the thermodynamic limit \cite{Wessel_Troyer_tri_XXZ,Boninsegni_Prokofev_tri_XXZ},
where $m$ is a positive number characterizing the strength of the
magnetization. However, although the effective spin-$1/2$ $XXZ$
model is a reasonable description of the original two-component Bose-Hubbard
model (\ref{eq:Hamiltonian}) in the strong coupling limit, one can
not exclude the influence of higher order terms neglected in it. As
a matter of fact, even within the spin-$1/2$ $XXZ$ model itself,
although $(\pm2m,\mp m,\mp m)$ is favored in the thermodynamic limit,
a metastable pattern $(\pm m,\mp m,0)$ is also found in quantum Monte
Carlo simulations on finite-sized lattices \cite{Wessel_Troyer_tri_XXZ,Boninsegni_Prokofev_tri_XXZ},
moreover a variational study shows that the energies corresponding
to these two patterns are very close to each other \cite{Sen_Moessner_tri_XXZ}.
In our simulations, we find that the system is close to a $(\pm m,\mp m,0)$
configuration (see Fig.~\ref{Flo:distr_magnetization}b), which could
indicate that the effective spin-$1/2$ $XXZ$ model plus higher order
corrections favors this pattern.

\subsubsection{Stripe SDW}

\begin{figure}
\includegraphics[width=3.3in]{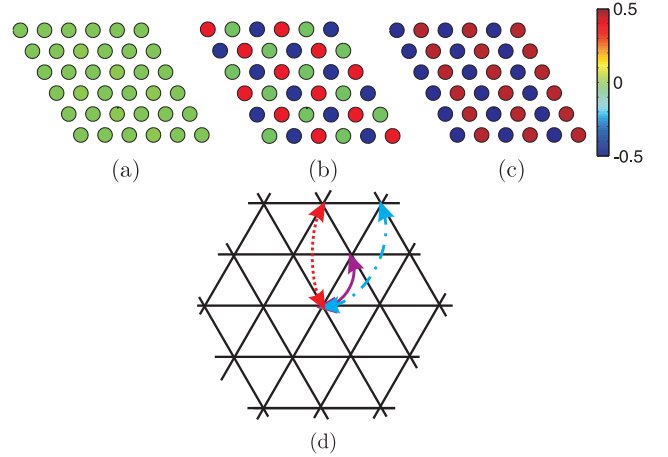}

\caption{(Color online) (a--c) $z$-magnetization distribution $\langle S_{i}^{z}\rangle$
for different Mott-insulating phases with quantum magnetic order.
(a) $xy$-ferromagnet for $9t_{a}/U=0.6$ and $9t_{b}/U=0.14$. (b)
3-sublattice SDW for $9t_{a}/U=0.7$ and $9t_{b}/U=0.1$. (c) Stripe
SDW for $9t_{a}/U=0.7$ and $9t_{b}/U=0.05$. In all the three cases,
we set $V_{a,b}/U=48$. (d) Schematic figure where nearest neighbor,
next-nearest neighbor, and next-next-nearest neighbor sites in a triangular
lattice are indicated by the purple solid line, red dashed line, and
blue dash-dotted line respectively.}

\label{Flo:distr_magnetization} 
\end{figure}

In the extremely asymmetric hopping region (area marked in red in
the phase diagram Fig.~\ref{Flo:pd_Vb_Vd_48U}a), we observe that
another type of SDW phase arises in the system. The $z$-magnetization
distribution is then characterized by a stripe pattern, shown in Fig.~\ref{Flo:distr_magnetization}c.
The transition from the stripe SDW phase to the homogeneous superfluid
is of first order, which leads to small coexistence regions (blue
areas in Fig.~\ref{Flo:pd_Vb_Vd_48U}a).

To understand the appearance of this stripe SDW, we notice that in
this regime the hopping amplitude of one species dominates. Without
loss of generality, in the following discussion we assume $t_{a}\gg t_{b}$,
which indicates that the $b$ species can be considered as almost
immobile scattering centers. If we assume that the effective spin-$1/2$
$XXZ$ model (\ref{eq:H_eff}) description still holds true in this
case (in principle the neglected higher order terms may be relevant),
the longitudinal exchange coupling $J_{z}$ is then much larger than
the transverse one $J_{\perp}$, hence the spin-$1/2$ $XXZ$ model
approximately reduces to an antiferromagnetic Ising model. From investigations
of the antiferromagnetic Ising model on a triangular lattice \cite{Ising_ant_ferro_tri},
we know that the stripe pattern of the magnetization is one of an
infinite number of degenerate ground state configurations. However,
here we observe no other patterns in our simulation except the stripe
configuration. This indicates that for quantitative insight one needs
to go beyond the spin-$1/2$ $XXZ$ model description, which we will
do in the following.

We notice that for extremely asymmetric hopping, together with the
condition that the intra-species interactions are much larger than
the inter-species interactions, i.e. $V_{a,b}\gg U$, the two-component
Bose-Hubbard model (\ref{eq:Hamiltonian}) can be simplified to a
bosonic Falicov-Kimball model with itinerant hard-core bosons \begin{equation}
H^{\mathrm{BFK}}=-\sum_{\langle i,j\rangle}(t_{a}a_{i}^{\dagger}a_{j}+\textnormal{h}.\textnormal{c}.)+U\sum_{i}n_{ai}n_{bi}-\sum_{i;\alpha=a,b}\mu_{\alpha}n_{\alpha i}\,.\label{eq:H_BFK}\end{equation}
 At zero temperature, integrating out the itinerant bosonic degree
of freedom under the assumption $t_{a}\ll U$, we end up with an effective
classical spin Hamiltonian\textcolor{blue}{{} }representing the density-density
interactions of the immobile $b$ species, which is accurate to the
order $O(t_{a}^{5}/U^{4})$ \cite{GS_FK_TRI} and reads

$ $

\begin{widetext} \begin{eqnarray}
H_{\mathrm{eff}}^{\mathrm{BFK}}(s;\mu_{\alpha}) & = & -\frac{1}{2}(\mu_{b}-\mu_{a})\sum_{i}s_{i}-\frac{1}{2}(\mu_{a}+\mu_{b}+U+\frac{3}{2}\frac{t_{a}^{3}}{U^{2}})N_{\mathrm{lat}}+\sum_{\langle i,j\rangle}\left[\frac{t_{a}^{2}}{4U}+\frac{t_{a}^{3}}{4U^{2}}-\frac{t_{a}^{4}}{8U^{3}}\right]s_{i}s_{j}\nonumber \\
 &  & +\sum_{\langle\langle i,j\rangle\rangle}\frac{5t_{a}^{4}}{16U^{3}}s_{i}s_{j}+\sum_{\langle\langle\langle i,j\rangle\rangle\rangle}\frac{t_{a}^{4}}{8U^{3}}s_{i}s_{j}-\sum_{P}\frac{t_{a}^{4}}{16U^{3}}(5+s_{P})\,,\label{eq:H_eff_BFK}\end{eqnarray}

\end{widetext} where $N_{\mathrm{lat}}$ is the number of lattice
sites, $\langle\langle i,j\rangle\rangle$ and $\langle\langle\langle i,j\rangle\rangle\rangle$
denote next-nearest neighbor and next-next-nearest neighbor sites
respectively (see Fig.~\ref{Flo:distr_magnetization}d). The Ising
pseudo spin $s_{i}$ is defined as $s_{i}\equiv(-1)^{n_{i}^{b}+1}$.
$P$ denotes the plaquettes made up of two triangles sharing an edge
and $s_{P}\equiv s_{P1}s_{P2}s_{P3}s_{P4}$ is the product of the
spins on all sites in the plaquette. Detailed studies of the model
(\ref{eq:H_eff_BFK}) in Ref. \cite{GS_FK_TRI} show that at the filling
$N_{a}=N_{b}=N/2$, the ground state of the above classical spin Hamiltonian
$H_{\mathrm{eff}}^{\mathrm{BFK}}(s;\mu_{\alpha})$ is non-degenerate
and has a stripe pattern configuration of the Ising pseudo spins similar
to that observed in our numerical BDMFT simulations. From (\ref{eq:H_eff_BFK})
we observe that in the leading order $O(t_{a}^{2}/U)$ the classical
spin model shows an Ising-type frustration on the triangular lattice,
however the frustration is removed by next-nearest neighbor and next-next-nearest
neighbor effective spin interactions which originate from higher-order
density fluctuations of the mobile particles (the $a$ species atoms
in this case).

\subsubsection{Weak charge-density wave (CDW) on top of SDW phase}

\begin{figure}
\includegraphics[width=3.3in]{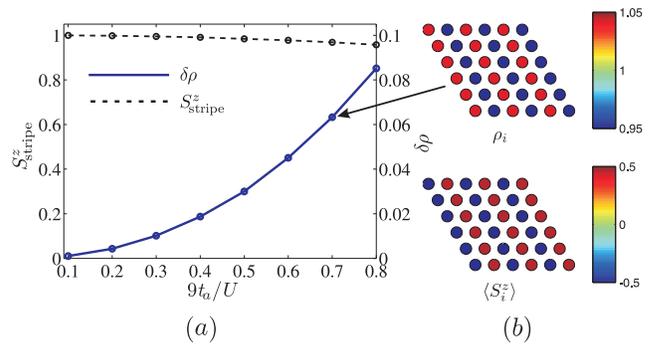}

\caption{(Color online) (a) In the stripe SDW phase region, we show the CDW
order (blue solid line) and SDW order (black dashed line) as a function
of $t_{a}$ when $9t_{b}/U=0.01$ is kept fixed. (b) The total density
($\rho_{i}$) (upper panel) and the $z$-magnetization ($\langle S_{i}^{z}\rangle$)
distribution (lower panel) for $V_{a,b}/U=48,\,9t_{a}/U=0.7$, and
$9t_{b}/U=0.01$ . }

\label{Flo:weak_CDW} 
\end{figure}

In the the SDW region we also observe a weak charge-density wave (CDW)
of the total density $\rho_{i}=\langle\sum_{\alpha=a,b}n_{\alpha i}\rangle$
forming on top of the SDW (see Fig.~\ref{Flo:weak_CDW}b). Specifically,
we found that for $t_{a}>t_{b}$, those lattice sites with a negative
$z$-magnetization have a larger total density $\rho_{i}$, while
for $t_{a}<t_{b}$, those sites with a positive $z$-magnetization
have the larger density.

The origin of this weak CDW can be easily understood by investigating
a simplified two-site two-component Bose-Hubbard model\begin{eqnarray}
H_{\mathrm{TT}} & = & -(t_{a}a_{L}^{\dagger}a_{R}+t_{b}b_{L}^{\dagger}b_{R}+\textnormal{h}.\textnormal{c}.)+U\sum_{i=L,R}n_{ai}n_{bi}\nonumber \\
 &  & +\frac{1}{2}\sum_{i=L,R;\alpha=a,b}V_{\alpha}n_{i\alpha}(n_{i\alpha}-1)\,.\label{eq:H_TT_1}\end{eqnarray}
 Since the onsite interaction is much larger than the kinetic energy,
we treat the hopping terms as perturbations. Since we are investigating
the CDW on top of the SDW, we assume a simple symmetry-broken state
$\left|\Psi_{G}^{(0)}\right\rangle =a_{L}^{\dagger}b_{R}^{\dagger}\left|0\right\rangle $
as the unperturbed ground state, which implies that the $L$ and $R$
sites have equal total density $(\rho_{L}=\rho_{R}=1)$ but opposite
$z$-magnetization ($\langle S_{L}^{z}\rangle=-\langle S_{R}^{z}\rangle=1/2$).
If we take into account the hopping terms to first order, the ground
state has the form \begin{equation}
\left|\Psi_{G}^{(1)}\right\rangle =\frac{Ua_{L}^{\dagger}b_{R}^{\dagger}\left|0\right\rangle +t_{b}a_{L}^{\dagger}b_{L}^{\dagger}\left|0\right\rangle +t_{a}a_{R}^{\dagger}b_{R}^{\dagger}\left|0\right\rangle }{\sqrt{U^{2}+t_{a}^{2}+t_{b}^{2}}},\end{equation}
 and the total density on the site $L$ and $R$ is \begin{eqnarray}
\rho_{L} & \thickapprox & 1-\delta,\\
\rho_{R} & \thickapprox & 1+\delta,\end{eqnarray}
 where $\delta=(t_{a}^{2}-t_{b}^{2})/(U^{2}+t_{a}^{2}+t_{b}^{2})$,
indicating that the weak CDW order on top of the SDW originates from
the asymmetry of the hopping amplitudes, i.e. the {}``lighter''
species can more easily delocalize to neighboring sites. Fig.~\ref{Flo:weak_CDW}
shows the amplitudes of the stripe SDW and the CDW orders as a function
of $t_{a}$ at fixed $t_{b}$. Those amplitudes are defined by $\delta\rho\equiv\rho_{+}-\rho_{-}$
and $S_{\textnormal{stripe}}^{z}\equiv S_{+}^{z}-S_{-}^{z}$ for CDW
and SDW order respectively, where $\rho_{+}$($\rho_{-}$) is the
total particle density per site on the stripe with higher (lower)
density and $S_{+}^{z}$($S_{-}^{z}$) is the $z$-magnetization per
site on the stripe with positive (negative) value.

More generally we can analyze the emergence of the CDW within a Ginzburg-Landau
framework. We assume spin density wave order at wavevectors $\mathbf{Q}$
and $-\mathbf{Q}$, and study its interaction with a putative CDW
order parameter. The lattice translation and Ising spin symmetries
of the problem determine the allowed coupling terms that affect the
CDW order, given in the following expansion of the free energy \begin{eqnarray}
F & = & \alpha_{s}|S_{\mathbf{Q}}^{z}|^{2}+\alpha_{\rho1}|\rho_{\mathbf{Q}}|^{2}+\alpha_{\rho2}|\rho_{2\mathbf{Q}}|^{2}\nonumber \\
 &  & +\beta_{1}(S_{\mathbf{Q}}^{z}S_{\mathbf{Q}}^{z}\rho_{-2\mathbf{Q}}+\textnormal{h}.\textnormal{c}.)\nonumber \\
 &  & +\beta_{2}|S_{\mathbf{Q}}^{z}|^{2}|\rho_{\mathbf{Q}}|^{2}+\beta_{3}\left[(S_{-\mathbf{Q}}^{z})^{2}(\rho_{\mathbf{Q}})^{2}+\textnormal{h}.\textnormal{c}.\right]\nonumber \\
 &  & +\gamma_{s}|S_{\mathbf{Q}}^{z}|^{4}+\gamma_{\rho1}|\rho_{\mathbf{Q}}|^{4}+\gamma_{\rho2}|\rho_{2\mathbf{Q}}|^{4},\label{eq:GL_F}\end{eqnarray}
where the $\alpha$'s, $\beta$'s, and $\gamma$'s are GL coefficients.
The crucial point is that symmetry allows a linear coupling of a charge
density wave at $\pm2\mathbf{Q}$ to the square of the SDW order parameter
(coupling $\beta_{1}$ above). Therefore a SDW order at $\pm\mathbf{Q}$
will necessarily produce CDW at $\mp2\mathbf{Q}$. The opposite is
not true, that is CDW order will not produce SDW in general. This
is because a quadratic-linear coupling to the SDW is prohibited by
the Ising spin symmetry. 

Let us first describe the implications of this fact in the case of
the 3-sublattice SDW with wavevector $\mathbf{Q}=(\pm4\pi/3,0)$.
In this case the CDW induced by the cubic coupling is of wavevector
$-2\mathbf{Q}=-\textnormal{sgn}(Q_{x})(4\pi,0)+\mathbf{Q}$, which
in the lattice is equivalent to $\mathbf{Q}$. This explains, from
a very general argument, why a CDW at $\mathbf{Q}$ must be produced
in this case. 

Let us move on to consider the case of the stripe SDW having a wavevector
$\mathbf{Q}=\pm\mathbf{K}/2$ with $\mathbf{K}$ being a reciprocal
lattice vector. The coupling to $\rho_{2\mathbf{Q}}$ is then equivalent
on the lattice to a coupling to the average density $\rho_{\mathbf{Q}=0}$
that can be absorbed into the chemical potential term. As a consequence,
the only non-trivial CDW's are the ones at $\pm\mathbf{Q}$ and induced
by the SDW order at $\pm\mathbf{Q}$ through the next order terms
quadratic in the CDW.

To briefly summarize, both 3-sublattice SDW and stripe SDW induce
CDW at the same wave vector, which is consistent with our simulations
(see Fig.~\ref{Flo:weak_CDW}b, which shows a stripe SDW with a large
SDW order induces a CDW at the same wave vector). However, in the
case of a stripe SDW the CDW is induced by coupling to $|\rho_{\mathbf{Q}}|^{2}$
rather than by linear coupling. It therefore requires a critical strength
of the stripe SDW to change the sign of the coefficient of $|\rho_{\mathbf{Q}}|^{2}$
and induce CDW order. 

\subsection{Superfluid and supersolid in the large-hopping region}

At sufficiently large hopping amplitudes, the ground state of the
system breaks $U(1)$ symmetry and develops superfluid long-range
order characterized by non-zero values of $\langle a\rangle$ or $\langle b\rangle$.
In the region where the hopping amplitudes $t_{a}$ and $t_{b}$ are
comparable to each other, we observe a second order phase transition
from $xy$-ferromagnetism to the homogeneous superfluid (HSF). On
the other hand, for highly asymmetric $t_{a}$ and $t_{b}$, increasing
the hopping amplitudes can first drive the system from the SDW into
a supersolid, which is characterized by coexisting superfluid and
CDW order. In the supersolid phase, the lighter species develops superfluid
order, i.e. $\langle\alpha\rangle\neq0$, while the heavier species
remains insulating, i.e. $\langle\bar{\alpha}\rangle=0$, where $\alpha(\bar{\alpha})$
denote the annihilation operator of the lighter/heavier species and
both species have CDW order in their density distribution respectively.
When the hopping amplitudes are further increased, a transition from
the supersolid to a homogeneous superfluid is observed.

\section{Conclusion}

We have investigated zero temperature quantum phases of Bose-Bose
mixtures in a triangular lattice using real-space Bosonic Dynamical
Mean Field Theory. A rich phase diagram including $xy$-ferromagnet,
spin-density wave, superfluid, and supersolid phases is found. In
the strong coupling regime, although an effective spin-$1/2$ $XXZ$
model gives qualitative insight, interesting phases beyond this effective
description are found: A stripe spin-density wave is identified for
highly asymmetric hopping, which originates from the interplay between
classical geometric frustration and higher order density fluctuations
of the lighter species. Moreover, on top of the spin-density wave,
due to asymmetric hopping amplitudes, the system shows a weak charge-density
wave in the total particle density distribution. 
\begin{acknowledgments}
L.~He acknowledges useful discussions with S. D. Huber, A. Sotnikov,
D. Cocks and I. Titvinidze and the hospitality of the Department of
Condensed Matter Physics, Weizmann Institute of Science, where parts
of this work were done. This work was supported by the Deutsche Forschungsgemeinschaft
via the DIP project HO 2407/5-1, Sonderforschungsbereich SFB/TR 49,
Forschergruppe FOR 801, ISF grant 1594/11 (E. A.) and by the China
Scholarship Fund (Y. L.). W. H. acknowledges the hospitality of KITP
during the final stages of this work, supported by the National Science
Foundation under Grant No. NSF PHY05-51164.

$ $\end{acknowledgments}

\end{document}